\documentclass[WPCF,manyauthors]{wpcfTemplate}

\usepackage{rotating}
\begin{document}%
%
%
\begin{titlepage}
%
\title{ CP and particle correlations under thermal influence }
\ShortTitle{ CP and particle correlations under thermal influence}
%
\author{G. Kozlov}
\author{
 JINR\\
}
\author{Email: kozlov@jinr.ru}
\ShortAuthor{XII$^{th}$ edition of WPCF}      
\begin{abstract}
We study the phase transition of nuclear (baryonic) matter in the model of one-dimensional random fluctuation walk. The stochastic fields (forces) influence intrinsic to Bose-Einstein correlations between two identical particles is a particular  ingridient of the model. The solution in terms of the characteristic functional equation contains the regular (homogenious)  and the singular non-analytical (non-homogenious) parts for baryonic matter (BM) and anomaly matter (AM), respectively.
The conditions to phase transition from BM to AM are found. 
\end{abstract}
\end{titlepage}
\section{Introduction}               
The critical behavior of baryonic matter (BM) and its transition to anomaly matter (AM) containing free fields of quarks and gluons at high temperature $T$ and baryon density  has a great interest, in particular, with the scanning of the QCD phase diagram to predict the Critical Point  (CP). The latter is characterized by the critical temperature $T_{c}$ and the high value of the critical baryon chemical potential $\mu_{c}$ (see, e.g., the refs. in [1]). The density $\rho_{q}$ of non-interacting quark gas with the temperature  $T$ depends on the number of flavors $N_{f}$ and baryon chemical potential $\mu$:
$$\frac{\rho_{q}}{T^{3}} \simeq 2\,N_{f}\,\frac{\mu}{3\,T}\left [ 1+ \frac{3}{\pi^{2}}\left (\frac{\mu}{3\,T}\right) ^{2}\right ].$$
In case of 3 light flavors one has $\rho_{q} \sim 6 (1 + 3/\pi^{2})T_{c}^{3}$ in the vicinity of CP if $\mu_{c}\simeq 3\,T_{c}$.
The large enough baryon density may be a source of spontaneous breaking of space parity of colliding particles (see, e.g., [2]). This breaking can be manifested [3] in the form of chiral magnetic effect [4] at non-central collisions of heavy ions. In addition, the local breaking of space parity (LBSP) of colliding nuclei may lead to anomaly yield of leptonic pairs [5,6]. LBSP may arise also in case of Bose-Einstein condensation (BEC) of light pseudoscalar particles, e.g., pions. 
In astrophysics (large space scales), BEC may be composed of axion-like objects localized inside the stars [7]. The appearance of pion condensate in hot and high density nuclear matter has not been discovered yet, and this might be one of the goals to the search programs of heavy ion machines operating at high enough baryon density.

It is known that the selfconsistent description of the critical behavior of the dynamical system in field theory is given by the renormalization group (RG) (see, e.g., [8-10] and the refs. therein). In RG approach to the field model the particular type of the phase transition is defined by the presence and the properties of the fixed points. As an example, the infra-red (IR)  attracted fixed point is usually relevant to the phase transition of the second kind as well as to the critical scaling. However, the fluxes (solutions) of RG may leave the physical domain containing the IR fixed point (IRFP). The fluxes may even go to infinity. Such a situtaion can be considered in terms of the phase transition of the first kind. In the neighborhood of an IRFP the strong coupling constant is 
$$\alpha^{IRFP}\simeq - 2\,\pi\frac{b_{0}}{b_{1}}, \,\,\, b_{0} = \frac{1}{3}\,(11\,N_{c} - 2\,N_{f}),$$
where $b_{0}$ and $b_{1}$ are the coefficients in the $\beta$ - function at some scale $\bar\mu$
$$\beta (\alpha) \equiv \bar\mu\frac{\partial\alpha}{\partial\bar\mu} = - \frac{b_{0}}{2\,\pi}\alpha^{2} - 
\frac{b_{1}}{(2\,\pi)^{2}}\alpha^{3} + ... $$
for $SU(N_{c})$ gauge theory. The scale of chiral-symmetry restoration and the associated deconfinement scale are of the order $\Lambda$ at which $\alpha (\bar\mu)$ crosses the critical value $\alpha_{c}$. In the walking gauge theory [11] accompanied by the solution to the linearized RG equation one has
$\beta (\alpha) \simeq \kappa \left (\alpha_{c} - \alpha^{IRFP}\right ) 
{\left (\bar\mu/\Lambda\right )}^{\kappa}, $
where $\alpha \simeq \alpha^{IRFP} - \left (\alpha^{IRFP} - \alpha_{c} \right ) 
{\left (\bar\mu/\Lambda\right )}^{\kappa} $. 
Here, $0 < \kappa\leq O(1)$, $0 < (\alpha^{IRFP} - \alpha _{c} ) << \alpha_{c}$, and $\beta$-function is small for conformal range $\bar\mu > \Lambda$. If $N_{f}$ decreases ($\alpha^{IRFP}$ increases) CP is characterized by $\alpha_{c} < \alpha^{IRFP}$ at which the spontaneous breaking of chiral symmetry is occurred and confinement does appear.

{\bf 2. Size of particle source}

One of useful and instructive approaches to CP study is through the spatial correlations of the final state particles.
To observe the size of the "hot" space composed of  particles produced in, e.g., heavy ion collisions, one needs to derive the theoretical formulas for two- or multi-particle distribution-correlation functions and the function of chaoticity of particles. The very popular hydrodynamical models can only describe the one-particle distribution, but never give any kind of particle number fluctuations and multi-particle correlations. We have already studied this subject in the frame of searching CP through the correlations between identical particles with Bose statistics, Bose-Einstein correlations (CBE), where the main object is the stochastic scale $L_{st}$ which defines the effective evolving size of the source of particles in hot excited matter [12]. 
$L_{st}$ depends on the temperature in the system (bath), the transverse momentum of two correlated particles, and feels the influence of random stochastic fields parametrised by the function of chaoticity strength $\nu$ which goes to zero when CP is approached. The stochastic scale enters CBE function $C_{2}(q;\lambda)$ in a simple form (for details, see [13]) 
\begin{equation}
\label{e1}
 C_{2}(q; \lambda)\simeq \eta (N) \left [1 + \lambda (\nu) \,e^{-q^{2}\, L_{st}^{2}}\right ]( 1 + \delta\cdot q + ...),
\end{equation}
where $\eta (N) = \langle N(N-1)\rangle /{\langle N\rangle}^{2}$ reflects "event-to-event" fluctuations to particle multitplicity $N$, $q$ is the relative momentum between two particles; $\lambda (\nu) = 1/(1 + \nu)^{2}$, where $0 <\nu < \infty $   [14]; $\delta$ is the measure of long-range effect that gives the distortion to $C_{2}$-function. Actually, $\delta\rightarrow 0$ at large enough space separation between two particles. On the other hand, in [15] there was reported an anti-correlation effect in which the measured correlation function $C_{2}$ attained values below the expected asymptotic minimum at unity, similarly to what had previously been observed in $e^{+}e^{-}$ collisions [16]. Phenomenological aspects of this dip-effect have been noted in [17].

The scaling form (\ref {e1}) is useful to predict the behavior of observables as well as to indicate the vicinity of CP. 
Note, that $ L_{st}\rightarrow\infty$ as $T\rightarrow T_{c}$ and  $\mu\rightarrow \mu_{c}$. One of the main questions is related to the observables. The latter may be associated, e.g.,  with the transverse momentum $k_{T} = \vert \vec p_{T_{1}} + \vec p_{T_{2}}\vert$ of two Bose particles with momenta $\vert \vec p_{T_{i}}\vert = \sqrt { \vec p_{x_{i}}^{2} +  \vec p_{y_{i}}^{2}}$, $i= 1,2$ at high enough $T$ [13]:
 \begin{equation}
\label{e2}
 k_{T}^{2} = \frac{1}{\nu (n)\, T^{3}\,L_{st}^5},\,\, \,\,n(\omega,\beta) = \frac{1}{e^{(\omega - \mu)\beta} -1},\,\,\,\,\, p_{\mu} = (\omega,\vec p).
\end{equation} 
The results in $e^{+}e^{-}$ collisions at LEP [18] as well as at the LHC energies by CMS [15,19], ALICE [20] and ATLAS [21] experiments reflected the decreasing size of particle source with increasing $k_{T}$ which is supported by (\ref{e2}) at finite $T$.
The coherence function $0 <\lambda < 1$ has been measured in many experiments, in low- and high energy accelerators (see, e.g., the refs in [22]. However, it is more important to study $\lambda$ theoretically as a function of $n$ at  the state with  the temperature $T$ through the transition from a fully coherent phase ($\lambda \simeq 0$) to the chaotic one ($\lambda \simeq 1$), where the critical behavior in the phase transition between BM and AM can occur. Thus, one can study the occurrence of Bose-Einstein condensation in thermal bath and two-particle momentum correlation as a function of an energy, (mean) particle multiplicity and the temperature of the system, distorted by the external fields (the influence of an environement). 

The light hadrons, e.g., pions, produced in heavy ion collisions, are characterized by temperature
$T_{\pi}$ close to pion mass. This is a pion condensate. However, the latter is a consequence of a hadronization process due to phase transition from free states of quarks and gluons (the vicinity of CP with $T_{c} > T_{\pi}$ ) through the mixed phase state composed of hadrons, quarks and gluons. Sometimes, this is called as the cross-over walk. 

To search the phase transition from BM to AM through CP we shall use the approach to random fluctuation walks with respect to chaoticity in correlations of identical particles.

{\bf 3. Random fluctuation walk}

We start with the simple model in which the phase transition between BM and AM modelling in one-dimensional $x$- oriented axis approach is governed by the probability $P(x;\bar\lambda, \mu_{c})$, where $-\infty < x <\infty$:
\begin{equation}
\label{e33}
 P(x;\bar\lambda, \mu_{c}) = p\,\sum_{j=0}^{\infty}\bar\lambda ^{j}\frac{1}{2}\sqrt {\frac{\pi}{t}} 
\left [e^{{-y^{2j}_{-}}/4t} +e^{{-y^{2j}_{+}}/4t} \right ],
\end{equation} 
where $\bar\lambda$ is the random fluctuation weigth,  $y^{j}_{\pm} = x\mu_{c} \pm a^{j}$, the running parameter $a = (\mu/\mu_{c}) > 1$; $t = l\mu_{c}$, $l$  is the lattice spacing. 
In the limit $l \rightarrow 0$ 
 ( $\mu_{c} \neq 0 $ ) one has (see also [23])
\begin{equation}
\label{e3}
 P(x;\bar\lambda, \mu_{c}) = p(\bar\lambda)\,\sum_{j=0}^{\infty}\bar\lambda ^{j}\,\pi \left [\delta (x\mu_{c} - a^{j}) + \delta (x\mu_{c} + a^{j})\right ],
\end{equation} 
where $p(\bar\lambda) = (1- \bar\lambda)/(2\pi)$ within the normalization condition $2 (p+\bar\lambda p + ...+ \bar\lambda ^{j} p + ...) =1$, $0 < \bar\lambda \leq 1$ 
The r.h.s. of (\ref{e3}) is the delta-shaped sequence of the r.h.s. (\ref{e33}) in the limit $t\rightarrow 0$.
The limit $\bar\lambda\rightarrow 1$  in (\ref{e3}) admits the broad behavior of $P(x;\bar\lambda, \mu_{c})$ that means the vicinity of CP is approached. Contrary to that, $P(x;\bar\lambda, \mu_{c})$ will be trivial if $\bar\lambda\rightarrow 0$: $P(x;\bar\lambda\rightarrow 0) \rightarrow 1/(2\pi)$. $\bar\lambda$ in (\ref{e3}) can be associated with the function $\lambda (\nu)$, introduced in (\ref{e1}), where [12] 
\begin{equation}
\label{e4}
\nu = \frac{1}{n\, k^{2}_{GL}}\, O \left (\frac{m^{2}_{\phi}}{m^{2}}\right ).
\end{equation}
Here, we have used the fact when CP is approaching, theory becomes conformal provided by the scalar field, the dilaton $\phi$, with the mass $m_{\phi}$. In (\ref{e4}) we use $m$ as a mass of light hadrons which are in the pattern of  CBE; $k_{GL} = m_{\phi}/m_{B}$  is the Ginzburg-Landau parameter which can differ the vacuum of type-I ( $k_{GL} <1$) from those of type-II ( $k_{GL} >1$) in dual Higgs-Abelian gauge model with dual (to non-abelian gluon field $A_{\alpha}(x)$) gauge field $B_{\alpha}(x)$ with the mass $m_{B}$ (for details see [12,17]. CP is characterized by $k_{GL}\rightarrow\infty$ because of infinite fluctuation length $\xi \sim m_{B}^{-1}$. The second central moment (dispersion) $m_{2}$ of (\ref{e3}) will be divergent at $\lambda\rightarrow 1$, thus the fluctuation length (the effective radius of the flux tube) $\xi\sim \sqrt {\vert m_{2}\vert}\rightarrow\infty$. In order to smooth the particularity (speciality) of (\ref{e3}) one can use its Fourier transformed form $G(k;\lambda,\mu_{c})$ which is nothing but the characteristic function. 
This provides the most convenient way to study an asymptotic behavior at large enough $x$ (or very sharp increasing of $L_{st}$ in terms of CBE function (\ref{e1})) corresponding to the vicinity of $k\rightarrow 0$. The finite values of $\xi$ will provide the analytical form of $G(k;\lambda,\mu_{c})$, however the very large fluctuation length gives the non-analytical behavior of $G(k;\lambda,\mu_{c})$ at $k\rightarrow 0$. The dual QCD vacuum (through $k_{GL}$) will infuence $\xi$ up to cross-over 
which is the unified process of phase transition between BM and AM. The function $G(k;\lambda,\mu_{c})$ has the form
\begin{equation}
\label{e5}
G(k;\lambda,\mu_{c}) = p(\lambda)\,\sum_{j=0}^{\infty} \lambda^{j}\,\cos\left ( \frac{k}{\mu_{c}}\,a^{j}\right ),
\end{equation}
accompanying by the properties: $2\pi\,G(0;\lambda) =1$, $G^{\prime} (0;\lambda) =0$. 
In some sense,  $G(k;\lambda)$ can be understand in terms of CBE function (\ref{e1}) in the form 
\begin{equation}
\label{e6}
G(k;\lambda,\mu_{c}) \sim \frac {const}{\bar C_{2}(q;\lambda) -1}, \,\,\, \bar C_{2}(q;\lambda) \equiv \frac{C_{2}(q;\lambda)}{\eta (n)},\,\, \,\,\,0 < \lambda < 1,
\end{equation}
where both sides for the first relation between $G$ and $\bar C_{2}$ in  (\ref{e6}) tend to infinity if $\lambda\rightarrow 1$. The fluctuation length $\xi$ is calculated through the even moments $m_{(2s)} (\lambda)$ of the order $2s$  
$$\xi^{2}_{(2s)} (\lambda) \sim m_{(2s)}(\lambda) = \frac{ \partial^{2s} G(k;\lambda,\mu_{c})}{\partial k^{2s}}\vert _{k=0}.$$
Looking through the result
\begin{equation}
\label{e7}
\vert \xi^{2}_{(2s)} (\lambda)\vert = p(\lambda)\,\sum_{j=0}^{\infty}{\left (\frac{a^{j}}{\mu_{c}}\right )}^{2s}\,\lambda ^{j},
\end{equation}
 one can realize that (\ref{e7}) does converge at  $(a/\mu_{c})^{2s}\,\lambda < 1$ only. In this case $\vert\xi^{2}\vert$ is finite
$$ \vert \xi^{2}_{(2s)} (\lambda)\vert \simeq p(\lambda)\,\mu^{-2s}_{c}\left ( 1+ a^{2s}\,\lambda\right ). $$
However, if $\mu \simeq \mu_{c}$ and $\lambda$ is close (but less) to 1 (this is a vicinity of CP), one finds the divergence of fluctuation length.

Actually, one can find the infinite number of divergent (singular) terms starting from (\ref{e5}), hence the latter does not suitable to describe an arbitrary phase of excited matter (wide values of $0 <\lambda  < 1$ and $\mu_{c}$, as well as small $k$ compared to $\mu_{c}$, $k\rightarrow 0$). To find the non-analytical part of (\ref{e5}) at $k =0$ let us start with functional scaling equation
(see also [23])
\begin{equation}
\label{e9}
G(k;\lambda, \mu_{c}) = G_{BM}(k;\lambda, \mu_{c}) + G_{AM}(ak;\lambda, \mu_{c}),
\end{equation}
which is the linear non-homogenious equation, where
\begin{equation}
\label{e10}
G_{BM}(k;\lambda, \mu_{c}) = p(\lambda)\,\cos(k/\mu_{c}),
\end{equation}
\begin{equation}
\label{e11}
G_{AM}(ak;\lambda, \mu_{c}) = \lambda\,G(ak;\lambda, \mu_{c}).
\end{equation}
(\ref{e10}) is the regular function at $k$ close to zero for all $0< \lambda < 1$ and it gives the contribution to BM. The phase with AM is provided by (\ref{e11}) at $ a^{-2} \leq \lambda < 1$. The second term in (\ref{e9}) disappears if $\lambda \rightarrow 0$, and the phase with BM does exist only. Here, we take into account that $ G(ak;\lambda \simeq 0, \mu_{c}) \simeq (1/2\pi)\, cos (k\mu/\mu^{2}_{c}) $ is finite. The same result at $\lambda \simeq 0$ is expected if one uses (\ref{e5}). 


The characteristic feature of the model is the fluctuation length (\ref{e7}) and its analytical and non-analytical behavior at all possible $0  < \lambda < 1$ and $\mu_{c}$. The solution for BM can be considered in the form of Taylor's expansion [23]
 \begin{equation}
\label{e12}
G_{BM} [k;\lambda (S), \mu_{c}] = 1 + \sum_{s=1}^{S-1} (-1)^{s}\,\frac {1}{(2\,s)!}\,\xi_{(2s)}^{2} [\lambda (S)]\, k^{2s},
\end{equation}
which is regular at $0  < \lambda < 1$ with finite $S$ ($S\rightarrow\infty$ as $\lambda\rightarrow 0$). The rest term for AM is singular for $\xi^{2}_{(2s)}$ at $s\geq S$:
\begin{equation}
\label{e13}
G_{AM} [k;\lambda (S), \mu_{c}] = 1 + \sum_{s=S}^{\infty} (i)^{2s}\,\frac {1}{(2\,s)!}\,\xi_{(2s)}^{2} [\lambda (S)]\, k^{2s}.
\end{equation}
Actually, for some values of $\lambda (S)$ at given $S$  with $k\rightarrow 0$ the rest (singular) term (\ref{e13}) can give the main contribution to (\ref{e9}) and, thus, shall define the asymptotic behavior of $G(k;\lambda, \mu_{c})$ at  $k\rightarrow 0$ and, consequently, the probablity $P(x;\lambda, \mu_{c}) $ in (\ref{e3}) at $x\rightarrow \infty$. The latter is in the correspondence with the infinite size of the particle source in terms of CBE at CP.

The special singular solution for AM may be presented in the $k$-power form [23]. Neglecting BM term (\ref{e10}) in (\ref{e9})  the infinite series (\ref{e13}) with the integer even powers $2s$ of $k$ is replaced by 
\begin{equation}
\label{e14}
G_{AM} [k;\lambda (S), \mu_{c}] = C[\lambda (S)]\,{\left \vert \frac{k}{\mu_{c}}\right \vert}^
{\alpha [\lambda (S)]}\,Q(\vert k\vert ),
\end{equation}
where $\alpha$ is restricted in some interval of $S$ to be defined later.
For any finite value $\lambda$ from an open interval (0,1) there will be the finite value $S(\lambda)$ from the semi-open interval [0,1), consequently, the inverse function $\lambda (S)$ does exist as well. The values of $\xi^{2}_{(2s)}(\lambda)$ are finite for $0 <\lambda < 1$ if $s =1,2,..., S-1$, and they will diverge if $s\geq S$. So, the series (\ref{e12}) is brought to the term $k^{2(S-1)}$.

Inserting (\ref{e14}) into the reduced homogenious equation (\ref{e9}) for AM, one finds
$$ \alpha[\lambda (S)] = \frac{\ln\left [\frac{Q(\vert k \vert)}{Q(a\vert k \vert)\cdot\lambda (S)}\right ]}{\ln a},\,\,\,\, a > 1. $$
Actually, $Q(a\vert k\vert)$ is the associated function of the 1st order to the measure $\gamma$ with the  proper function of the operator of dilatation transformation $u$, $u\,Q(\vert k\vert) = Q(a\vert k\vert)$. For any $a \geq 1$ the function $Q(a\vert k\vert)$ obeys the following condition [24]
$$Q(a\vert k\vert) = a^{\gamma}\,Q (\vert k \vert) + a^{\gamma}\,\ln (a)\cdot\, Q_{0}(\vert k\vert),$$
where $Q_{0}$ is the homogenious function of the measure $\gamma$. 
For $a$ close to unity (from above) one finds
$$ \alpha[\lambda (S)] = -\left [\frac{\ln\lambda (S)}{\ln a} + \gamma\right ]. $$
The singularity of (\ref{e14}) is supported by the unequality 
$$ 2(S+1) - \gamma\ln a \geq \alpha (S) > 2S - \gamma\ln a. $$
The coefficient $C [\lambda (S)]$ in (\ref{e14}) is 
$$ C[\lambda (S)] = \frac{i^{\alpha -2}}{\Gamma(\alpha-1)}\,\xi^{2}_{(\alpha -2)}(\lambda). $$
The solution for $Q(\vert k\vert)$ in (\ref{e14}) obeying  the functional dilatation equation
$ u\,Q(\vert k\vert) \simeq Q(a\cdot\vert k\vert)$
contains the smoothly changing function which will be, e.g., the logarithmic-periodic function with the period $\log(a) > 0$. If one replaces $\vert k\vert\rightarrow \log (\vert k\vert )$, $a\cdot \vert k\vert\rightarrow \log (a) +\log (\vert k\vert )$, the solution of corresponding log-periodic equation is the infinite series
$$ Q(\vert k\vert) \sim \frac{1}{\log (a)}\,\sum_{m=0}^{\infty} b_{m}\,\cos\left [2\pi m \frac{\log (\vert k\vert)}{\log (a)} + \varphi_{m}\right ], $$
 where $b_{m}$ and $\varphi_{m}$ are coefficients and the phases, respectively, which are not important for the model considered here. 
Thus, the solution for AM looks like
\begin{equation}
\label{e20}
G_{AM}(k;\lambda, \mu_{c}) \sim \frac{i^{\alpha-2}}{\Gamma (\alpha -1)}\xi^{2}_{(\alpha -2)} (\lambda) {\left \vert\frac{k}{\mu_{c}}\right \vert}^{\alpha} \frac{1}{\log (a)}\,\sum_{m=0}^{\infty} b_{m}\cos\left [2\pi m \frac{\log (\vert k\vert)}{\log (a)} + \varphi_{m}\right ],
\end{equation}
which is divergent if $\lambda\rightarrow 1$ (because of $\xi^{2}$), and $a\rightarrow 1$ from above (because of $\log (a)$ in the denominator of (\ref{e20}). The AM solution disappears if $\alpha (S)/S \rightarrow 2$ as $S\rightarrow \infty$ at $\lambda (S)\rightarrow 0$.
If we are close to AM phase the Ginzburg-Landau coefficient $k_{GL}$ is given by the unequality 
$$ k^{2}_{GL} \geq \frac{1}{n}\,\frac{2}{\left (\mu/\mu_{c}\right )^{2} -1} $$
and the cross-over  does occur when $k_{GL}\rightarrow\infty$ at $\mu = \mu_{c}$ with no dependence on the particles occupation number $n$.

{\bf 4. Conclusions }

The theoretical search for CP, in paricular, the cross-over between BM and AM, is studied in one-dimensional model of random fluctuation walk accompanied by the known model of CBE. The main points here are the random fluctuation weight $\bar\lambda$ in (\ref{e3}), the even moments of $G$ or the fluctuation length squared $\xi^{2}$, and the running parameter $a >1$. $\bar\lambda$ is the stochastic random  function $\lambda (\nu)$ which defines the strength of correlations between two identical objects with Bose-statistics. This depends very strongly of the vacuum properties characterized by Ginzburg-Landau parameter $k_{GL}$. It is shown that the solution is given in terms of the regular and the singular parts corresponding to BM and AM, respectively. The smooth $\lambda$ provides BM, however, if $\lambda\rightarrow 1$ (very strong influence of chaoticity) the asymptotic behavior at $k\rightarrow 0$ (or the  probabiilty $P(x)$ for $x\rightarrow\infty$) is approached. The latter corresponds to infinite size $L_{st}$ of a particle source .

\end{document}